\documentstyle[12pt]{article}
\newcommand{\ret}{\nonumber \\}

\newcommand{\Section}[1]%
{\section{#1}\setcounter{equation}{0}%
\setcounter{theorem}{0}}
\newtheorem{theorem}{Theorem}
\newtheorem{lemma}[theorem]{Lemma}
\newtheorem{coro}[theorem]{Corollary}

\newtheorem{definition}[theorem]{Definition}
\newtheorem{assumption}[theorem]{Assumption}

\newenvironment{proof}[1]%
{\par\noindent{\em #1:\ }}%
{~\rule{2mm}{2mm}\par\bigskip}
\begin{document}
\newpage\thispagestyle{empty}
{\topskip 2cm
\begin{center}
{\Large\bf Spectral Gap and Exponential Decay of Correlations\\} 
\bigskip\bigskip\bigskip\bigskip
{\large Matthew B. Hastings$^1$\\}
\bigskip
{\large Tohru Koma$^2$\\}
\bigskip\bigskip
\noindent
$^1$ 
{\small \it Center for Nonlinear Studies and Theoretical Division,
Los Alamos National Laboratory, Los Alamos, 
New Mexico 87545, USA}\\
\noindent
{\small\tt e-mail: hastings@lanl.gov}\\
\medskip
\noindent
$^2$ 
{\small \it Department of Physics, Gakushuin University, 
Mejiro, Toshima-ku, Tokyo 171-8588, JAPAN}\\
\noindent
{\small\tt e-mail: tohru.koma@gakushuin.ac.jp}

\end{center}
\bigskip\bigskip\bigskip
\noindent
We study the relation between the spectral gap above 
the ground state and the decay of the correlations in the ground state  
in quantum spin and fermion systems with short-range interactions 
on a wide class of lattices.
We prove that, if two observables anticommute with each other at
large distance, then the nonvanishing spectral gap implies 
exponential decay of the corresponding correlation. 
When two observables commute with each other at
large distance, the connected correlation function
decays exponentially under the gap assumption.  If the observables
behave as a vector under the U(1) rotation 
of a global symmetry of the system, we use previous results on
the large distance decay of the correlation function to show the
stronger statement that the
correlation function itself, rather than just the connected correlation
function, decays exponentially under the gap assumption 
on a lattice with a certain self-similarity in (fractal) dimensions $D<2$. 
In particular, if the system is translationally invariant in one of the 
spatial directions, then this self-similarity condition is automatically 
satisfied. 
We also treat systems with long-range, power-law decaying interactions.   
\par
\vfil}\newpage
\Section{Introduction}

In non-relativistic quantum many-body systems, a folk theorem states that 
a nonvanishing spectral gap above the ground state implies 
exponentially decaying correlations in the ground state. 
Perhaps this has been the most popular folk theorem 
in this field since Haldane \cite{Haldane} predicted a ``massive phase" in 
low dimensional, isotropic quantum systems.   
Quite recently, this statement was partially proved \cite{Koma} 
for quantum lattice systems with a global U(1) symmetry 
in (fractal) dimensions $D<2$. More precisely, 
a bound which decays to zero at large distance was obtained  
for correlation functions whose observables behave as a vector 
under the U(1)-rotation. Unfortunately, the bound is weaker than 
the expected exponential decay.  
On the other hand, exponential clustering of the correlations 
was also proved recently \cite{Hastings1,Hastings2} for quantum many-body lattice systems  
under the gap assumption. This is a non-relativistic version of 
Fredenhagen's theorem \cite{Fredenhagen,NachSim} of relativistic quantum 
field theory. Clearly the following
natural question arises: can this clustering property 
be combined with the above bound for the decay of the correlations to yield
the tighter, exponentially decaying bound for the correlation functions
themselves, rather than just for the connected correlation functions?  
We emphasize that these are different statements; given clustering,
the decay of the correlation functions requires also that certain
matrix elements vanish in the ground state sector.

In this paper, we address this problem and reexamine the above folk theorem 
by relying on the exponential clustering of the correlations.  
Our first step is to provide a rigorous proof of the exponential clustering.
We extend the previous results in this case to
treat long-range interactions including both power-law and exponentially 
decaying interactions.   
In the former case, all the upper bounds for the correlations become 
power-law bounds.

We then prove that ground state correlation functions of 
observables which transform as vectors under a U(1) symmetry
decay exponentially or
with a power law, depending on the form of  the interaction,
given an additional assumption on a certain self-similarity.
In particular, if the system is translationally invariant in one of the 
spatial directions,  
this self-similarity condition is automatically satisfied. 
Therefore the corresponding correlation functions decay 
exponentially for translationally invariant 
systems on one-dimensional regular lattices.  
As a byproduct, we also prove that, if two observables anticommute with each other 
at large distance, then the corresponding correlation in 
the ground state decays exponentially under the gap assumption 
for a wide class of lattice fermion systems with exponentially decaying interactions in 
any dimensions. In this case, we do not need any other assumption except for 
those on the interactions and the spectral gap. 

This paper is organized as follows: In the next section, we give 
the precise definitions of the models, and describe our main results.  
In Section~\ref{sec:clustering}, we prove the clustering of generic correlation
functions under the gap assumption, 
and obtain the upper decaying bound for the fermionic correlations. 
The decay of the bosonic correlations are treated in Section~\ref{sec:vanishing}.  
Appendix~\ref{LRbound} is devoted to the proof of the Lieb-Robinson bound 
for the group velocity of the information propagation in the models 
with a long-range interaction decaying by power law. 

\Section{Models and main results}
\label{Sec:main}

We consider quantum systems on generic lattices \cite{Fractal}. 
Let $\Lambda_s$ be a set of the sites, $x,y,z,w,\ldots$, and 
$\Lambda_b$ a set of the bonds, i.e., pairs of sites, 
$\{x,y\},\{z,w\},\ldots$. We call the pair, $\Lambda:=(\Lambda_s,\Lambda_b)$, 
the lattice. If a sequence of sites, $x_0,x_1,x_2,\ldots,x_n$, 
satisfies $\{x_{j-1},x_j\}\in\Lambda_b$ for $j=1,2,\ldots,n$, then 
we say that the path, $\{x_0,x_1,x_2,\ldots,x_n\}$, has length $n$ and 
connects $x_0$ to $x_n$.  
We denote by ${\rm dist}(x,y)$ the graph-theoretic distance 
which is defined to be the shortest path length that one needs to connect $x$ to $y$. 
We denote by $|X|$ the cardinality of the finite set $X$. 
The Hamiltonian $H_\Lambda$ is defined on the tensor product 
$\bigotimes_{x\in\Lambda_s}{\cal H}_x$ of a finite dimensional 
Hilbert space ${\cal H}_x$ at each site $x$. 
We assume $\sup_{\Lambda_s}\sup_x{\rm dim}\>{\cal H}_x\le N<\infty$. 
For a lattice fermion system, we consider the Fock space. 

Consider the Hamiltonian of the form, 
\begin{equation}
H_\Lambda=\sum_{X\subset\Lambda_s}h_X, 
\end{equation}
where $h_X$ is the local Hamiltonian of the compact support $X$. 
We consider both power-law and exponentially decaying interactions $h_X$. 

For the power-law decaying interactions $h_X$, 
we require the following conditions: 

\begin{assumption}
\label{assumption:powerdecay}
The interaction $h_X$ satisfies  
\begin{equation}
\sum_{X\ni x,y}
\Vert h_X\Vert\le\frac{\lambda_0}{[1+{\rm dist}(x,y)]^\eta}
\label{powerdecayhX} 
\end{equation}
with positive constants, $\lambda_0$ and $\eta$, and the lattice $\Lambda$ equipped with 
the metric satisfies  
\begin{equation}
\sum_{z\in{\Lambda_s}}\frac{1}{[1+{\rm dist}(x,z)]^\eta}
\times\frac{1}{[1+{\rm dist}(z,y)]^\eta}\le
\frac{p_0}{[1+{\rm dist}(x,y)]^\eta}
\label{Propaineq}
\end{equation}
with a positive constant $p_0$.   
\end{assumption}

\noindent
{\bf Remark:}
If  
\begin{equation}
\sup_{\Lambda_s}\>\sup_x\sum_{y\in{\Lambda_s}}
\frac{1}{[1+{\rm dist}(x,y)]^\eta}<\infty,  
\label{decaysumcondition}
\end{equation}
then the inequality (\ref{Propaineq}) holds as follows: 
\begin{eqnarray}
& &\sum_{z\in{\Lambda_s}}\frac{1}{[1+{\rm dist}(x,z)]^\eta}
\times\frac{1}{[1+{\rm dist}(z,y)]^\eta}\ret
&=&\frac{1}{[1+{\rm dist}(x,y)]^\eta}
\sum_{z\in{\Lambda_s}}\frac{[1+{\rm dist}(x,y)]^\eta}{[1+{\rm dist}(x,z)]^\eta
[1+{\rm dist}(z,y)]^\eta}\ret
&\le&\frac{1}{[1+{\rm dist}(x,y)]^\eta}
\sum_{z\in{\Lambda_s}}2^\eta\frac{[1+{\rm dist}(x,z)]^\eta+
[1+{\rm dist}(z,y)]^\eta}{[1+{\rm dist}(x,z)]^\eta
[1+{\rm dist}(z,y)]^\eta}\ret
&\le&\frac{1}{[1+{\rm dist}(x,y)]^\eta}
\sum_{z\in{\Lambda_s}}2^\eta\left\{\frac{1}{[1+{\rm dist}(x,z)]^\eta}
+\frac{1}{[1+{\rm dist}(z,y)]^\eta}\right\}, 
\end{eqnarray}
where we have used the inequality, 
$[1+{\rm dist}(x,y)]^\eta\le 2^\eta([1+{\rm dist}(x,z)]^\eta
+[1+{\rm dist}(z,y)]^\eta)$. 
{From} the assumption (\ref{powerdecayhX}) and the condition (\ref{decaysumcondition}), 
one has  
\begin{equation}
\sup_x \sum_{X\ni x}\Vert h_X\Vert|X|\le s_0<\infty,
\end{equation}
where $s_0$ is a positive constant 
which is independent of the volume of $|\Lambda_s|$.

Instead of these conditions, we can also require:  

\begin{assumption}
\label{assumption:powerdecaysum}
The interaction $h_X$ satisfies
\begin{equation}
\sup_x\sum_{X\ni x}\Vert h_X\Vert|X|[1+{\rm diam}(X)]^\eta\le s_1<\infty,
\label{s1bound}
\end{equation}
where $\eta$ is a positive constant, ${\rm diam}(X)$ is 
the diameter of the set $X$, i.e., ${\rm diam}(X)=\max\{{\rm dist}(x,y)|\>x,y\in X\}$, 
and $s_1$ is a positive constant which is independent of the volume of $|\Lambda_s|$.   
\end{assumption}

For exponentially decaying interactions $h_X$, we require one of the 
following two assumptions: 

\begin{assumption}
\label{assumption:expodecay}
There exists a positive $\eta$ satisfying the condition (\ref{decaysumcondition}). 
The interaction $h_X$ satisfies   
\begin{equation}
\sum_{X\ni x,y}
\Vert h_X\Vert\le\lambda_0\exp[-(\mu+\varepsilon)\>{\rm dist}(x,y)]
\label{expodecayhX} 
\end{equation}
with some positive constants, $\lambda_0,\mu$ and $\varepsilon$. 
\end{assumption}

\noindent
{\bf Remark:} From the conditions, we have 
\begin{equation}
\exp[-(\mu+\varepsilon)\>{\rm dist}(x,y)]\le 
\frac{\lambda_0'\exp[-\mu\>{\rm dist}(x,y)]}{[1+{\rm dist}(x,y)]^\eta}
\label{expepspoly}
\end{equation}
with a positive constant $\lambda_0'$, and 
\begin{equation}
\sum_{z\in{\Lambda_s}}\frac{\exp[-\mu\>{\rm dist}(x,z)]}{[1+{\rm dist}(x,z)]^\eta}
\times\frac{\exp[-\mu\>{\rm dist}(z,y)]}{[1+{\rm dist}(z,y)]^\eta}\le
\frac{p_0\exp[-\mu\>{\rm dist}(x,y)]}{[1+{\rm dist}(x,y)]^\eta}
\label{expotwoprod}
\end{equation}
with a positive constant $p_0$ in the same way as in the preceding remark.
\bigskip

\begin{assumption}
\label{assumption:expodecaysum}
The interaction $h_X$ satisfies    
\begin{equation}
\sup_x\sum_{X\ni x}
\Vert h_X\Vert|X|\exp[{\mu\>{\rm diam}(X)}]\le s_1<\infty,
\end{equation}
where $\mu$ is a positive constant, and 
$s_1$ is a positive constant which is independent of the volume of $|\Lambda_s|$.
\end{assumption}

\noindent
{\bf Remark:} This assumption is milder than that in \cite{NachSim}
by the absence of the factor $N^{2|X|}$ in the summand.  
\bigskip

Further we assume the existence of a ``uniform gap" above the ground
state sector of the Hamiltonian $H_\Lambda$. 
The precise definition of the ``uniform gap" is:  

\begin{definition}{\bf (Uniform gap):} 
\label{definition}
We say that there is a uniform gap above the ground state sector
if the spectrum $\sigma(H_\Lambda)$ of the Hamiltonian $H_\Lambda$ 
satisfies the following conditions: 
The ground state of the Hamiltonian $H_\Lambda$ is 
$q$-fold (quasi)degenerate in the sense that there are $q$ eigenvalues, 
$E_{0,1},\ldots, E_{0,q}$, in the ground state sector
at the bottom of the spectrum of $H_\Lambda$ such that 
\begin{equation}
\Delta{\cal E}:=\max_{\mu,\mu'}\{|E_{0,\mu}-E_{0,\mu'}|\}\rightarrow 0\quad\mbox{as }\
|\Lambda_s|\rightarrow\infty.
\label{defDeltacalE} 
\end{equation}
Further the distance between the spectrum, $\{E_{0,1},\ldots, E_{0,q}\}$,   
of the ground state and the rest of the spectrum is larger than 
a positive constant $\Delta E$ which is independent of the volume $|\Lambda_s|$. 
Namely there is a spectral gap $\Delta E$ above the ground state sector. 
\end{definition}

Let $A_X,B_Y$ be observables with the support $X,Y\subset\Lambda_s$, respectively.  
We say that the pair of two observables, $A_X$ and $B_Y$, is fermionic 
if they satisfy the anicommutation relation, $\{A_X,B_Y\}=0$ 
for $X\cap Y=\emptyset$. 
If they satisfy the commutation relation, then we call the pair bosonic.  

Define the ground-state expectation as 
\begin{equation}
\left\langle\cdots\right\rangle_{0,\Lambda}:=\frac{1}{q}{\rm Tr}\ (\cdots)P_{0,\Lambda},
\label{GSexpect}
\end{equation}
where  $P_{0,\Lambda}$ is the projection onto the ground state sector. 
For the infinite volume, 
\begin{equation}
\left\langle\cdots\right\rangle_0:={\rm weak}^\ast\mbox{-}
\lim_{|\Lambda_s|\uparrow\infty}
\left\langle\cdots\right\rangle_{0,\Lambda},
\end{equation}
where we take a suitable subsequence 
of finite lattices $\Lambda$ going to the infinite volume so that 
the expectation converges to a linear functional for 
a set of quasilocal observables.
Although the ground-state expectation thus constructed depends on 
the subsequence of the lattices and on the observables, our results below hold 
for any ground-state expectation thus constructed.  
Further, we denote by 
\begin{equation}
\omega(\cdots):={\rm weak}^\ast\mbox{-}\lim_{|\Lambda_s|\uparrow\infty}
\left\langle\Phi_\Lambda,(\cdots)\Phi_\Lambda\right\rangle
\label{omega}
\end{equation}
the ground-state expectation in the infinite volume 
for a normalized vector $\Phi_\Lambda$ in the sector of the ground state 
for finite lattice $\Lambda$.    

\begin{theorem}{\bf (Clustering of fermionic correlations):}
\label{clusterboundfermionic} 
Let $A_X,B_Y$ be fermionic observables with a compact support. 
Assume that there exists a uniform spectral gap $\Delta E>0$ above the 
ground state sector in the spectrum of the Hamiltonian 
$H_\Lambda$ in the sense of Definition~\ref{definition}. 
Let $\omega$ be a ground-state expectation (\ref{omega}) 
in the infinite volume limit. Then the following bound is valid:    
\begin{eqnarray}
& &\left|\omega(A_XB_Y)
-\frac{1}{2}\left[\omega(A_XP_0B_Y)-\omega(B_YP_0A_X)\right]\right|\ret
&\le&{\rm Const.}\times\cases{[1+{\rm dist}(X,Y)]^{-{\tilde\eta}}, 
                             & for power-law decaying $h_X$;\cr
                    \exp[-{\tilde \mu}\>{\rm dist}(X,Y)], 
                             & for exponentially decaying $h_X$,\cr}
\label{clsfermionic}
\end{eqnarray}
where $P_0$ is the projection onto the sector of the infinite-volume 
ground state,\footnote{$\omega(\cdots P_0\cdots)$ is also defined as 
a bilinear functional for a set of quasilocal observables in the weak$^\ast$ limit.} and   
\begin{equation}
{\tilde \eta}=\frac{\eta}{1+2v_\eta/\Delta E}\quad\mbox{and}\quad
{\tilde \mu}=\frac{\mu}{1+2v_\mu/\Delta E}.
\label{tildeetamu}
\end{equation}
Here $v_\eta$ and $v_\mu$ are, respectively, an increasing function of $\eta$ and $\mu$, 
and give an upper bound of the group velocity of the information propagation.  
\end{theorem} 

\noindent
{\bf Remark:} Clearly there exists a maximum $\mu_{\rm max}$ such that 
the bound (\ref{expodecayhX}) holds for any $\mu\le \mu_{\rm max}$. 
Combining this observation with (\ref{tildeetamu}), there exists a maximum 
${\tilde\mu}=\max_{\mu\le\mu_{\rm max}}\{\mu/(1+2v_\mu/\Delta E)\}$ which 
gives the optimal decay bound. When the interaction $h_X$ is of finite 
range, one can take any large $\mu$. But the upper bound $v_\mu$ of 
the group velocity exponentially increases as $\mu$ increases 
because $v_\mu$ depends on $\lambda_0$ of (\ref{expodecayhX}). 
In consequence, a finite ${\tilde \mu}$ gives the optimal bound.  

Formally applying the identity, 
$\langle A_XP_0B_Y\rangle_0=\langle B_YP_0A_X\rangle_0$, 
for the bound (\ref{clsfermionic}), 
we have the following decay bound for the correlation:\footnote{See 
Section~\ref{sec:clustering} for details.}  

\begin{coro}
Let $A_X,B_Y$ be fermionic observables with a compact support. 
Assume that there exists a uniform spectral gap $\Delta E>0$ above the 
ground state sector in the spectrum of the Hamiltonian 
$H_\Lambda$ in the sense of Definition~\ref{definition}. 
Then the following bound is valid:    
\begin{equation}
\left|\left\langle A_XB_Y\right\rangle_0\right|
\le{\rm Const.}\times\cases{[1+{\rm dist}(X,Y)]^{-{\tilde\eta}}, 
                             & for power-law decaying $h_X$;\cr
                    \exp[-{\tilde \mu}\>{\rm dist}(X,Y)], 
                             & for exponentially decaying $h_X$,\cr}
\label{decayboundfermionic} 
\end{equation}
in the infinite volume limit, where 
$\tilde \eta,\tilde \mu$ are as defined above. 
\end{coro} 

\begin{theorem}{\bf (Clustering of bosonic correlations):} 
Let $A_X,B_Y$ be bosonic observables with a compact support. 
Assume that there exists a uniform spectral gap $\Delta E>0$ above the 
ground state sector in the spectrum of the Hamiltonian 
$H_\Lambda$ in the sense of Definition~\ref{definition}. 
Let $\omega$ be a ground-state expectation (\ref{omega}) 
in the infinite volume limit. Then the following bound is valid:    
\begin{eqnarray}
& &\left|\omega(A_XB_Y)
-\frac{1}{2}\left[\omega(A_XP_0B_Y)+\omega(B_YP_0A_X)\right]\right|\ret
&\le&{\rm Const.}\times\cases{[1+{\rm dist}(X,Y)]^{-{\tilde\eta}}, 
                             & for power-law decaying $h_X$;\cr
                    \exp[-{\tilde \mu}\>{\rm dist}(X,Y)], 
                             & for exponentially decaying $h_X$,\cr}
\end{eqnarray}
where $\tilde \eta,\tilde \mu$ are as defined above.
\label{clusterboundbosonic} 
\end{theorem}

\noindent
{\bf Remark:} Theorem~\ref{clusterboundbosonic} is a clustering bound for
the connected correlation functions.  We now make some additional definitions
that will enable us, in certain cases, to prove the decay of
$\left[\omega(A_XP_0B_Y)+\omega(B_YP_0A_X)\right]/2$ so that
Theorem~\ref{clusterboundbosonic} can be replaced with a stronger bound
below, Theorem~\ref{theorem:bosonic}.

\begin{definition}{\bf (Self-similarity):}
\label{SSassumption}
Write $m=q^2$ with the degeneracy $q$ of the ground state sector. 
We say that the system has self-similarity if the following 
conditions are satisfied: 
For any observable $A$ of compact support and any given large $L>0$, 
there exist transformations, $R_1,R_2,\ldots,R_m$, and 
observables, $B^{(1)},B^{(2)},\ldots,B^{(m)}$, such that  
the Hamiltonian $H_\Lambda$ is invariant under the transformations, 
i.e., $R_j(H_\Lambda)=H_\Lambda$ for any lattice $\Lambda$ with
sufficiently large $|\Lambda_s|$,  
and that the observables satisfy the following conditions:
\begin{equation}
B^{(j)}=R_j(A)\quad\mbox{and}\quad
\left(B^{(j)}\right)^\dagger=R_j(A^\dagger)\quad\mbox{for }\ j=1,2,\ldots,m,
\end{equation}
\begin{equation}
{\rm dist}({\rm supp}\ A,{\rm supp}\ B^{(j)})\ge L\quad\mbox{for }\ 
j=1,2,\ldots,m,
\label{distABL}
\end{equation}
and 
\begin{equation}
{\rm dist}({\rm supp}\ B^{(j)},{\rm supp}\ B^{(k)})\ge L\quad\mbox{for }\ j\ne k.
\label{distBBL}
\end{equation}
\end{definition}
\medskip

In Section~\ref{sec:vanishing}, we will discuss other conditions similar to 
this self-similarity condition.

\begin{theorem}
\label{theorem:bosonic}
Assume that the degeneracy $q$ of the ground state sector
of the Hamiltonian $H_\Lambda$ is finite in the infinite volume limit, 
and that there exists a uniform spectral gap $\Delta E>0$ above the 
ground state sector in the spectrum of the Hamiltonian 
$H_\Lambda$ in the sense of Definition~\ref{definition}. 
Further assume that the system has self-similarity in the sense of 
Definition~\ref{SSassumption}, and that there exists a subset ${\cal A}_b^s$ of 
bosonic observables with a compact support such that 
$R_j({\cal A}_b^s)\subset{\cal A}_b^s=({\cal A}_b^s)^\dagger$ for $j=1,2,\ldots,m$, 
and that $\left\langle A_X'B_Y'\right\rangle_0\rightarrow 0$ as 
${\rm dist}(X,Y)\rightarrow\infty$ 
for any pair of bosonic observables, $A_X',B_Y'\in{\cal A}_b^s$. 
Let $\omega$ be a ground-state expectation (\ref{omega}) 
in the infinite volume limit, and let $A_X,B_Y$ be a pair of bosonic observables 
satisfying $A_X\in{\cal A}_b^s$. Then the following bound is valid:    
\begin{equation}
\left|\omega(A_XB_Y)\right|
\le{\rm Const.}\times\cases{[1+{\rm dist}(X,Y)]^{-{\tilde \eta}}, 
                             & for power-law decaying $h_X$;\cr
                    \exp[-{\tilde \mu}\>{\rm dist}(X,Y)], 
                             & for exponentially decaying $h_X$,\cr}
\label{decaycorrboson} 
\end{equation}
where $\tilde \eta,\tilde \mu$ are as defined above. 
\end{theorem} 

\noindent
{\bf Remark:} 1. If the finite system is translationally invariant in one of 
the spatial directions with a periodic boundary condition, then 
the self-similarity condition of Definition~\ref{SSassumption} is automatically satisfied 
by taking the translation as the transformation $R_j$. Thus we do not 
need an additional assumption for such systems.\\
\noindent
2. Theorem~\ref{theorem:bosonic} can be extended to a system 
having infinite degeneracy 
of the ground state sector in the infinite volume limit
if the degeneracy for finite volume is sufficiently small 
compared to the volume of the system. See Theorem~\ref{theorem:vanishfinite} in  
Section~\ref{sec:vanishing} for details.
\bigskip

In order to apply this theorem, we need to be able to show that
$\left\langle A_XB_Y\right\rangle_0\rightarrow 0$ as 
${\rm dist}(X,Y)\rightarrow\infty$ 
in the
infinite volume for any pair of bosonic observables, $A_X,B_Y\in{\cal A}_b^s$. 
However, this was proven \cite{Koma} 
for quantum spin or fermion 
systems with a global U(1) symmetry on a class of lattices with 
(fractal) dimension $D<2$ as defined in (\ref{dimensionD})
below,
so long as the 
observables behave as a vector 
under the U(1) rotation.

We first define the dimension for these lattices.
The ``sphere", $S_r(x)$, centered at $x\in \Lambda_s$ with the radius $r$ is 
defined as  
\begin{equation}
S_r(x):=\{y\in\Lambda_s|{\rm dist}(y,x)=r\}. 
\end{equation}
Assume that there exists 
a ``(fractal) dimension" $D\ge 1$ of the lattice $\Lambda$ such that 
the number $|S_r(x)|$ of the sites in the sphere satisfies
\begin{equation}
\sup_{x\in\Lambda_s}|S_r(x)|\le C_0r^{D-1}
\label{dimensionD}
\end{equation}
with some positive constant $C_0$. 
This class of the lattices is the same as in~\cite{KT2}.

Consider spin or fermion systems with a global U(1) symmetry 
on the lattice $\Lambda$ with (fractal) dimension $1\le D<2$, 
and require the existence of a uniform gap above the ground state sector
of the Hamiltonian $H_\Lambda$ in the sense of 
Definition~\ref{definition}. 
Although the method of \cite{Koma} can be 
applied to a wide class of such systems, we consider only two
important examples, the Heisenberg and the Hubbard models.
We take the set ${\cal A}_b^s$ to be the bosonic observables 
which behave as a vector under the U(1) rotation. 
In the rest of this section we use the results of \cite{Koma} to show as in
(\ref{dcybound1},\ref{dcybound2}) that the correlation function for this
class of observables in these models does decay to zero
as ${\rm dist}(X,Y)\rightarrow\infty$.  The bounds (\ref{dcybound1},\ref{dcybound2})
provide only a slow bound on the decay.  However, this slow
bound on the decay suffices, in conjunction with
the self-similarity condition of Definition~\ref{SSassumption} 
to apply Theorem~\ref{theorem:bosonic}.
{\em Thus, under the self-similarity assumption as well as the gap
assumption, all the upper bounds below (\ref{dcybound1},\ref{dcybound2})
are replaced with exponentially 
decaying bounds by Theorem~\ref{theorem:bosonic}. In particular, 
a system with a translational invariance automatically satisfies the 
self-similarity condition as mentioned above. Therefore the corresponding 
correlations show exponential decay for translationally 
invariant systems on one-dimensional regular lattices.}
\bigskip

\noindent
{\it XXZ Heisenberg model:} The Hamiltonian $H_\Lambda$ is given by 
\begin{equation}
H_\Lambda=H_\Lambda^{XY}+V_\Lambda(\{S_x^{(3)}\})
\label{hamspin}
\end{equation}
with
\begin{equation}
H_\Lambda^{XY}=2\sum_{\{x,y\}\in\Lambda_b}J_{x,y}^{\rm XY}
\left[S_x^{(1)}S_y^{(1)}+S_x^{(2)}S_y^{(2)}\right],
\label{hamXY}
\end{equation}
where ${\bf S}_x=(S_x^{(1)},S_x^{(2)},S_x^{(3)})$ is the spin operator 
at the site $x\in\Lambda_s$ with the spin $S=1/2,1,3/2,\ldots$, and 
$J_{x,y}^{\rm XY}$ are real coupling constants; 
$V_\Lambda(\{S_x^{(3)}\})$ is a real function of the $z$-components, 
$\{S_x^{(3)}\}_{x\in\Lambda_s}$, 
of the spins.   
For simplicity, we take 
\begin{equation}
V_\Lambda(\{S_x^{(3)}\})=\sum_{\{x,y\}\in\Lambda_b}J_{x,y}^{\rm Z}
S_x^{(3)}S_y^{(3)}
\end{equation}
with real coupling constants $J_{x,y}^{\rm Z}$. 
Assume that there are positive constants, $J_{\rm max}^{\rm XY}$ and 
$J_{\rm max}^{\rm Z}$, which satisfy $|J_{x,y}^{\rm XY}|\le J_{\rm max}^{\rm XY}$  
and $|J_{x,y}^{\rm Z}|\le J_{\rm max}^{\rm Z}$ for any bond $\{x,y\}\in\Lambda_b$. 

Consider the transverse spin-spin correlation, 
$\left\langle S_x^+S_y^-\right\rangle_0$, where $S_x^\pm:=S_x^{(1)}\pm i S_x^{(2)}$.

\begin{theorem}
Assume that the fractal dimension $D$ of (\ref{dimensionD}) satisfies 
$1\le D<2$, and that there exists a uniform spectral gap $\Delta E>0$ above the 
ground state sector in the spectrum of the Hamiltonian 
$H_\Lambda$ of (\ref{hamspin}) in the sense of Definition~\ref{definition}.  
Then there exists a positive constant $\gamma$ such that 
the transverse spin-spin correlation satisfies the bound,  
\begin{equation}
\left|\left\langle S_x^+S_y^-\right\rangle_0\right|\le
{\rm Const.}\exp\left[-\gamma\{{\rm dist}(x,y)\}^{1-D/2}\right], 
\label{dcybound1}
\end{equation}
in the thermodynamic limit $|\Lambda_s|\rightarrow\infty$. 
\end{theorem}

The proof is given in \cite{Koma}, and we remark that the result can be extended to 
more complicated correlations such as 
the multispin correlation, 
$\left\langle S_{x_1}^+\cdots S_{x_j}^+S_{y_1}^-\cdots S_{y_j}^-\right\rangle_0$. 
If the system satisfies the self-similarity condition of 
Definition~\ref{SSassumption},
then the upper bound, (\ref{dcybound1}),  
can be replaced with a stronger exponentially decaying one 
by Theorem~\ref{theorem:bosonic}.
\medskip

\noindent
{\it Hubbard model \cite{KT1,MR}:} 
The Hamiltonian on the lattice $\Lambda$ is given by 
\begin{equation}
H_\Lambda=-\sum_{\{x,y\}\in\Lambda_b}\sum_{\alpha=\uparrow,\downarrow}
\left(t_{x,y}c_{x,\alpha}^\dagger c_{y,\alpha}+t_{x,y}^\ast 
c_{y,\alpha}^\dagger c_{x,\alpha}
\right)+V(\{n_{x,\alpha}\})+\sum_{x\in\Lambda_s}{\bf B}_x\cdot{\bf S}_x,
\label{hamelectron}
\end{equation}
where $c_{x,\alpha}^\dagger, c_{x,\alpha}$ are, respectively, 
the electron creation and annihilation operators with the $z$ component 
of the spin $\mu=\uparrow,\downarrow$, $n_{x,\alpha}=c_{x,\alpha}^\dagger 
c_{x,\alpha}$ is the corresponding number operator, and 
${\bf S}_x=(S_x^{(1)},S_x^{(2)},S_x^{(3)})$ are the spin operator given by 
$S_x^{(a)}=\sum_{\alpha,\beta=\uparrow,\downarrow}c_{x,\alpha}^\dagger 
\sigma_{\alpha,\beta}^{(a)}
c_{x,\beta}$ with the Pauli spin matrix $(\sigma_{\alpha,\beta}^{(a)})$ for $a=1,2,3$; 
$t_{i,j}\in{\bf C}$ are the hopping amplitude, $V(\{n_{x,\alpha}\})$ is a real 
function of the number operators, and ${\bf B}_x=(B_x^{(1)},B_x^{(2)},B_x^{(3)})\in{\bf R}^3$ 
are local magnetic fields. Assume that the interaction $V(\{n_{x,\alpha}\})$ is  
of finite range in the sense of the graph theoretic distance.   

\begin{theorem}
Assume that the fractal dimension $D$ of (\ref{dimensionD}) satisfies 
$1\le D<2$, and that there exists a uniform spectral gap $\Delta E>0$ above the 
ground state sector in the spectrum of the Hamiltonian 
$H_\Lambda$ of (\ref{hamelectron}) in the sense of Definition~\ref{definition}. 
Then the following bound is valid:  
\begin{equation}
\left|\left\langle c_{x,\uparrow}^\dagger c_{x,\downarrow}^\dagger 
c_{y,\uparrow}c_{y,\downarrow}\right\rangle_0\right|\ret
\le{\rm Const.}\exp\left[-\gamma\{{\rm dist}(x,y)\}^{1-D/2}\right] 
\label{dcybound2} 
\end{equation}
with some constant $\gamma$ 
in the thermodynamic limit $|\Lambda_s|\rightarrow\infty$.
If the local magnetic field has the form ${\bf B}_x=(0,0,B_x)$, then 
we further have 
\begin{equation}
\left|\left\langle S_x^+ S_y^-\right\rangle_0\right|\le
{\rm Const.}\exp\left[-\gamma'\{{\rm dist}(x,y)\}^{1-D/2}\right] 
\label{dcybound3}  
\end{equation} 
with some constant $\gamma'$. 
\end{theorem}

The proof is given in \cite{Koma}. 
Clearly the Hamiltonian $H_\Lambda$ of (\ref{hamelectron}) commutes with the total number 
operator ${\cal N}_\Lambda=\sum_{x\in\Lambda_s}\sum_{\mu=\uparrow,\downarrow}n_{x,\mu}$ 
for a finite volume $|\Lambda_s|<\infty$. 
We denote by $H_{\Lambda,N}$ the restriction of $H_\Lambda$ onto 
the eigenspace of ${\cal N}_\Lambda$ with the eigenvalue $N$. 
Let $P_{0,\Lambda,N}$ be the projection onto the ground state sector
of $H_{\Lambda,N}$, and we denote the ground-state expectation by 
\begin{equation}
\left\langle\cdots\right\rangle_{0,\nu}={\rm weak}^\ast\mbox{-}
\lim_{|\Lambda_s|\uparrow\infty}\frac{1}{q_N}
{\rm Tr}\ (\cdots)P_{0,\Lambda,N},  
\end{equation}
where $q_N$ is the degeneracy of the ground state, and $\nu$ is the limit of 
the filling factor $N/|\Lambda_s|$ of the electrons. 
Since the operators $S_x^\pm$ do not connect the sectors 
with the different eigenvalues $N$, we have 

\begin{theorem}
Assume that the fractal dimension $D$ of (\ref{dimensionD}) satisfies 
$1\le D<2$, and that there exists a uniform spectral gap $\Delta E>0$ above the 
ground state sector in the spectrum of the Hamiltonian 
$H_{\Lambda,N}$ in the sense of Definition~\ref{definition}. 
Then the following bound is valid for the filling factor $\nu$ of the electrons:      
\begin{equation}
\left|\left\langle S_x^+ S_y^-\right\rangle_{0,\nu}\right|\le
{\rm Const.}\exp\left[-\gamma'\{{\rm dist}(x,y)\}^{1-D/2}\right] 
\label{dcybound4}  
\end{equation} 
with some constant $\gamma'$ in the infinite volume limit.
\end{theorem}

The proof is given in \cite{Koma}. 
If the system satisfies the self-similarity condition of Definition~\ref{SSassumption},
then these three upper bounds, (\ref{dcybound2}), (\ref{dcybound3}) 
and (\ref{dcybound4}),  
can be replaced with a stronger exponentially decaying one 
by Theorem~\ref{theorem:bosonic}.

\Section{Clustering of correlations}
\label{sec:clustering}

In order to prove the power-law and the exponential clustering, 
Theorems~\ref{clusterboundfermionic} and \ref{clusterboundbosonic}, we follow 
the method \cite{Hastings1}. The key tools of the proof 
are Lemma~\ref{lemma:Hastings} below and 
the Lieb-Robinson bound \cite{NachSim,LR} 
for the group velocity of the information propagation. 
The sketch of the proof is that 
the static correlation function can be derived from the time-dependent 
correlation function by the lemma, and the large-distance behavior of 
the time-dependent correlation function is estimated by the Lieb-Robinson
bound.  As a byproduct, we obtain  
the decay bound (\ref{decayboundfermionic}) for fermionic observables. 

Consider first the case of the bosonic observables. 
Let $A_X,B_Y$ be bosonic observables with compact supports $X,Y\subset\Lambda_s$, 
respectively, and let $A_X(t)=e^{itH_\Lambda}A_Xe^{-itH_\Lambda}$, 
where $t\in{\bf R}$ and $H_\Lambda$ is the Hamiltonian for finite 
volume. Let $\Phi$ be a normalized vector in the ground state sector. 
The ground state expectation of the commutator is written as  
\begin{eqnarray}
\left\langle\Phi,[A_X(t),B_Y]\Phi\right\rangle
&=&\left\langle\Phi,A_X(t)(1-P_{0,\Lambda})B_Y\Phi\right\rangle
-\left\langle\Phi,B_Y(1-P_{0,\Lambda})A_X(t)\Phi\right\rangle\ret
&+&\left\langle\Phi,A_X(t)P_{0,\Lambda}B_Y\Phi\right\rangle
-\left\langle\Phi,B_YP_{0,\Lambda}A_X(t)\Phi\right\rangle. 
\label{ABcommuexpect}
\end{eqnarray}
In terms of the ground state vectors $\Phi_{0,\nu}, \nu=1,2,\ldots,q$, 
with the energy eigenvalues, $E_{0,\nu}$, and the excited state vectors 
$\Phi_n$ with $E_n, n=1,2,\ldots$, one has 
\begin{equation}
\left\langle\Phi,A_X(t)(1-P_{0,\Lambda})B_Y\Phi\right\rangle
=\sum_{\nu,\nu'}\sum_{n\ne 0}a_\nu^\ast a_{\nu'}
\left\langle\Phi_{0,\nu},A_X\Phi_n\right\rangle
\left\langle\Phi_n,B_Y\Phi_{0,\nu'}\right\rangle
e^{-it(E_n-E_{0,\nu})},
\label{negativepart}
\end{equation}
\begin{equation}
\left\langle\Phi,B_Y(1-P_{0,\Lambda})A_X(t)\Phi\right\rangle
=\sum_{\nu,\nu'}\sum_{n\ne 0}a_\nu^\ast a_{\nu'}
\left\langle\Phi_{0,\nu},B_Y\Phi_n\right\rangle
\left\langle\Phi_n,A_X\Phi_{0,\nu'}\right\rangle e^{it(E_n-E_{0,\nu'})},
\label{positivepart}
\end{equation}
\begin{equation}
\left\langle\Phi,A_X(t)P_{0,\Lambda}B_Y\Phi\right\rangle
=\sum_{\nu,\nu'}\sum_{\mu}a_\nu^\ast a_{\nu'}
\left\langle\Phi_{0,\nu},A_X\Phi_{0,\mu}\right\rangle
\left\langle\Phi_{0,\mu},B_Y\Phi_{0,\nu'}\right\rangle
e^{-it(E_{0,\mu}-E_{0,\nu})} 
\end{equation}
and 
\begin{equation}
\left\langle\Phi,B_YP_{0,\Lambda}A_X(t)\Phi\right\rangle
=\sum_{\nu,\nu'}\sum_{\mu}a_\nu^\ast a_{\nu'}
\left\langle\Phi_{0,\nu},B_Y\Phi_{0,\mu}\right\rangle
\left\langle\Phi_{0,\mu},A_X\Phi_{0,\nu'}\right\rangle
e^{it(E_{0,\mu}-E_{0,\nu'})}, 
\end{equation}
where we have written 
\begin{equation}
\Phi=\sum_{\nu=1}^q a_\nu\Phi_{0,\nu}.
\end{equation}

In order to get the bound for $\left\langle\Phi,A_X(t=0)B_Y\Phi\right\rangle$, 
we want to extract only the ``negative frequency 
part" (\ref{negativepart}) from 
the time-dependent correlation functions (\ref{ABcommuexpect}).
For this purpose, we use the following lemma \cite{Hastings1}:

\begin{lemma}
\label{lemma:Hastings}
Let $E\in{\bf R}$, and $\alpha>0$. Then 
\begin{eqnarray}
\lim_{T\uparrow\infty}\lim_{\epsilon\downarrow 0}
\frac{i}{2\pi}\int_{-T}^T\frac{e^{-iEt}e^{-\alpha t^2}}{t+i\epsilon}dt
&=&\frac{1}{2\pi}\sqrt{\frac{\pi}{\alpha}}
\int_{-\infty}^0d\omega\exp[-(\omega+E)^2/(4\alpha)]\ret
&=&\cases{1+{\cal O}(\exp[-\Delta E^2/(4\alpha)]) & for $E\ge \Delta E$;\cr
{\cal O}(\exp[-\Delta E^2/(4\alpha)]) & for $E\le -\Delta E$.\cr}
\label{limitIE}
\end{eqnarray}
\end{lemma}

\begin{proof}{Proof}
Write 
\begin{equation}
I(E)=\frac{i}{2\pi}\int_{-T}^T\frac{e^{-iEt}e^{-\alpha t^2}}{t+i\epsilon}dt.
\end{equation}
Using the Fourier transformation,
\begin{equation}
e^{-iEt}e^{-\alpha t^2}=\frac{1}{2\pi}\sqrt{\frac{\pi}{\alpha}}
\int_{-\infty}^\infty\exp[-(\omega+E)^2/(4\alpha)]e^{i\omega t}d\omega,
\label{FTRexpress}
\end{equation}
we decompose the integral $I(E)$ into three parts as  
\begin{equation}
I(E)=I_-(E)+I_0(E)+I_+(E),
\label{Ftdecompo}
\end{equation}
where 
\begin{equation}
I_-(E)=\frac{i}{2\pi}\frac{1}{2\pi}\sqrt{\frac{\pi}{\alpha}}
\int_{-T}^T\ dt\frac{1}{t+i\epsilon}\int_{-\infty}^{-\Delta\omega}d\omega
\exp[-(\omega+E)^2/(4\alpha)]e^{i\omega t},
\end{equation}
\begin{equation}
I_0(E)=\frac{i}{2\pi}\frac{1}{2\pi}\sqrt{\frac{\pi}{\alpha}}
\int_{-T}^T\ dt\frac{1}{t+i\epsilon}\int_{-\Delta\omega}^{\Delta\omega}d\omega
\exp[-(\omega+E)^2/(4\alpha)]e^{i\omega t},
\end{equation}
and
\begin{equation}
I_+(E)=\frac{i}{2\pi}\frac{1}{2\pi}\sqrt{\frac{\pi}{\alpha}}
\int_{-T}^T\ dt\frac{1}{t+i\epsilon}\int_{\Delta\omega}^\infty d\omega
\exp[-(\omega+E)^2/(4\alpha)]e^{i\omega t},
\label{I+}
\end{equation}
where we choose $\Delta\omega=bT^{-1/2}$ with some positive constant $b$. 

First let us estimate $I_0(E)$. Note that 
\begin{equation}
\frac{1}{t+i\epsilon}=\frac{t}{t^2+\epsilon^2}
-\frac{i\epsilon}{t^2+\epsilon^2}.
\end{equation}
Using this identity, one has 
\begin{equation}
I_0(E)=\frac{i}{2\pi}\frac{1}{2\pi}\sqrt{\frac{\pi}{\alpha}}
\int_{-\Delta\omega}^{\Delta\omega}d\omega
\exp[-(\omega+E)^2/(4\alpha)]
\int_{-T}^T\ dt\left[\frac{t\sin\omega t}{t^2+\epsilon^2}
-\frac{i\epsilon\cos\omega t}{t^2+\epsilon^2}\right],
\end{equation}
where we have interchanged the order of the double integral by relying on 
$|t|\le T<\infty$. Since the integral about $t$ can be bounded by 
some constant, one obtains  
\begin{equation}
|I_0(E)|\le {\rm Const.}\times \alpha^{-1/2}\Delta\omega
\le {\rm Const.}\times \alpha^{-1/2}T^{-1/2}.
\label{F0bound}
\end{equation}
Therefore the corresponding contribution is vanishing 
in the limit $T\uparrow\infty$. 

Note that 
\begin{equation}
\frac{i}{2\pi}\int_{-T}^T\ dt\frac{e^{i\omega t}}{t+i\epsilon}
=\cases{{\cal O}(\omega^{-1}T^{-1}) & for $\omega>0$;\cr
e^{\epsilon\omega}+{\cal O}(\omega^{-1}T^{-1}) & for $\omega<0$.\cr}
\end{equation}
Using this, the function $I_+(E)$ of (\ref{I+}) can be evaluated as  
\begin{equation}
|I_+(E)|\le{\rm Const.}\times T^{-1/2}.
\label{F+bound} 
\end{equation}
This is also vanishing in the limit.  

Thus it is enough to consider only the integral $I_-(E)$. 
In the same way as the above, one has 
\begin{equation}
I_-(E)=\frac{1}{2\pi}\sqrt{\frac{\pi}{\alpha}}
\int_{-\infty}^{-\Delta\omega}d\omega\exp[-(\omega+E)^2/(4\alpha)]
e^{\epsilon\omega}+{\cal O}(T^{-1/2}).
\end{equation}
Since $e^{\epsilon\omega}\le 1$ for $\omega<0$, one has 
\begin{equation}
\lim_{T\uparrow\infty}\lim_{\epsilon\downarrow 0}
I_-(E)=\frac{1}{2\pi}\sqrt{\frac{\pi}{\alpha}}
\int_{-\infty}^0d\omega\exp[-(\omega+E)^2/(4\alpha)]. 
\end{equation}
Note that, for $E\le -\Delta E$, 
\begin{equation}
\frac{1}{2\pi}\sqrt{\frac{\pi}{\alpha}}
\int_{-\infty}^0d\omega\exp[-(\omega+E)^2/(4\alpha)]
\le\frac{1}{2}\exp[-\Delta E^2/(4\alpha)],  
\end{equation}
and, for $E\ge\Delta E$, 
\begin{eqnarray}
\frac{1}{2\pi}\sqrt{\frac{\pi}{\alpha}}
\int_{-\infty}^0d\omega\exp[-(\omega+E)^2/(4\alpha)]
&=&\frac{1}{2\pi}\sqrt{\frac{\pi}{\alpha}}
\int_{-\infty}^{\infty}d\omega\exp[-(\omega+E)^2/(4\alpha)]\ret
&-&\frac{1}{2\pi}\sqrt{\frac{\pi}{\alpha}}
\int_0^{\infty}d\omega\exp[-(\omega+E)^2/(4\alpha)]\ret
&=&1+{\cal O}(\exp[-\Delta E^2/(4\alpha)]).
\end{eqnarray}
Clearly these imply (\ref{limitIE}).
\end{proof}

{From} Lemma~\ref{lemma:Hastings} and the expression (\ref{ABcommuexpect})
of the correlation function with (\ref{negativepart}) and (\ref{positivepart}), 
one has 
\begin{eqnarray}
& &\lim_{T\uparrow\infty}\lim_{\epsilon\downarrow 0}\frac{i}{2\pi}\int_{-T}^T\ dt 
\frac{1}{t+i\epsilon}
\left\langle\Phi,[A_X(t),B_Y]\Phi\right\rangle e^{-\alpha t^2}\ret
&=&\left\langle\Phi,A_X(1-P_{0,\Lambda})B_Y\Phi\right\rangle+
{\cal O}(\exp[-\Delta E^2/(4\alpha)])\ret
&+&\lim_{T\uparrow\infty}\lim_{\epsilon\downarrow 0}\frac{i}{2\pi}\int_{-T}^T\ dt 
\frac{1}{t+i\epsilon}
\left[\left\langle\Phi,A_X(t)P_{0,\Lambda}B_Y\Phi\right\rangle
-\left\langle\Phi,B_YP_{0,\Lambda}A_X(t)\Phi\right\rangle\right]e^{-\alpha t^2}
\label{extractTrcorr} 
\end{eqnarray}
for finite volume. 

In the following, we treat only the power-law decaying 
interaction $h_X$ because one can treat the exponentially decaying  
interactions in the same way. See also refs.~\cite{Hastings1,Hastings2} 
in which the exponential clustering of 
the correlations is proved for finite-range interactions under the gap assumption 
along the same line as below. 

In order to estimate the left-hand side, 
we recall the Lieb-Robinson estimate (\ref{LRboundpower}) 
in Appendix~\ref{LRbound}, 
\begin{equation}
\left\Vert\frac{1}{t}[A_X(t),B_Y]\right\Vert\le {\rm Const.}\times 
\frac{1}{(1+r)^\eta}\frac{e^{v|t|}-1}{|t|}, 
\end{equation}
for $r>0$, where we have written $r={\rm dist}(X,Y)$.   
Using this estimate, the integral can be evaluated as  
\begin{eqnarray}
& &\left|\int_{-T}^T\ dt 
\frac{\left\langle\Phi,[A_X(t),B_Y]\Phi\right\rangle}{t+i\epsilon}e^{-\alpha t^2}\right|
\ret
&\le&\left|\int_{|t|\le c\ell}\ dt 
\frac{\left\langle\Phi,[A_X(t),B_Y]\Phi\right\rangle}{t+i\epsilon}e^{-\alpha t^2}\right|
+\left|\int_{|t|> c\ell}\ dt 
\frac{\left\langle\Phi,[A_X(t),B_Y]\Phi\right\rangle}{t+i\epsilon}e^{-\alpha t^2}\right|\ret
&\le&{\rm Const.}\times \frac{1}{(1+r)^{\eta-cv}} 
+\frac{{\rm Const.}}{\sqrt{\alpha}\ell}\exp[-\alpha c^2\ell^2],
\label{LRboundresult}
\end{eqnarray}
where $c$ is a positive, small parameter, and $\ell=\log(1+r)$, and 
we have used 
\begin{equation}
\int_{|t|\le c\ell}\frac{e^{v|t|}-1}{|t|}dt\le 2e^{cv\ell}.
\end{equation}

In order to estimate the integral in the right-hand side of (\ref{extractTrcorr}),  
we consider the matrix element 
$\left\langle\Phi_{0,\nu},A_X(t)P_{0,\Lambda}B_Y\Phi_{0,\nu'}\right\rangle$ 
because the other matrix elements in the ground state 
can be treated in the same way. 
Using Lemma~\ref{lemma:Hastings}, one has 
\begin{eqnarray}
& &\lim_{T\uparrow\infty}\lim_{\epsilon\downarrow 0}
\frac{i}{2\pi}\int_{-T}^T dt 
\frac{1}{t+i\epsilon}
\left\langle\Phi_{0,\nu},A_X(t)P_{0,\Lambda}B_Y\Phi_{0,\nu'}\right\rangle
e^{-\alpha t^2}\ret
&=&\sum_{\mu=1}^q\left\langle\Phi_{0,\nu},A_X\Phi_{0,\mu}\right\rangle
\left\langle\Phi_{0,\mu},B_Y\Phi_{0,\nu'}\right\rangle
\frac{1}{2\pi}\sqrt{\frac{\pi}{\alpha}}\int_{-\infty}^0
d\omega\exp[-(\omega+\Delta{\cal E}_{\mu,\nu})^2/(4\alpha)],\ret
\end{eqnarray}
where $\Delta{\cal E}_{\mu,\nu}=E_{0,\mu}-E_{0,\nu}$. 
Using the assumption (\ref{defDeltacalE}) and the dominated convergence theorem, 
we have that, for any given $\varepsilon>0$, there exists a sufficiently large 
volume of the lattice $\Lambda_s$ such that 
\begin{equation}
\left|\lim_{T\uparrow\infty}\lim_{\epsilon\downarrow 0}
\frac{i}{2\pi}\int_{-T}^T dt 
\frac{e^{-\alpha t^2}}{t+i\epsilon}
\left\langle\Phi_{0,\nu},A_X(t)P_{0,\Lambda}B_Y\Phi_{0,\nu'}\right\rangle
-\frac{1}{2}\left\langle\Phi_{0,\nu},A_XP_{0,\Lambda}B_Y\Phi_{0,\nu'}\right\rangle\right|
<\varepsilon.  
\end{equation}
Combining this observation, (\ref{extractTrcorr}) and (\ref{LRboundresult}), 
and choosing $\alpha=\Delta E/(2c\ell)$, one obtains 
\begin{eqnarray}
& &\left|\omega(A_XB_Y)
-\frac{1}{2}\left[\omega(A_XP_0B_Y)+\omega(B_YP_0A_X)\right]\right|\ret
&\le&{\rm Const.}\times\frac{1}{(1+r)^{\eta-cv}}
+{\rm Const.}\times \exp\left[-\frac{c\Delta E}{2}\ell\right] 
\end{eqnarray}
in the infinite volume limit, where the ground-state expectation $\omega$ 
is given by (\ref{omega}). 
Choosing $c=\eta/(v+\Delta E/2)$, we have 
\begin{equation}
\left|\omega(A_XB_Y)
-\frac{1}{2}\left[\omega(A_XP_0B_Y)+\omega(B_YP_0A_X)\right]\right|
\le\frac{{\rm Const.}}{[1+{\rm dist}(X,Y)]^{\tilde\eta}},
\label{expdecaytruncated}
\end{equation}
with ${\tilde\eta}=\eta/(1+2v/\Delta E)$. 
In the same way, we have  
\begin{equation}
\left|\omega(A_XB_Y)
-\frac{1}{2}\left[\omega(A_XP_0B_Y)+\omega(B_YP_0A_X)\right]\right|
\le{\rm Const.}\times\exp[-{\tilde \mu}\>{\rm dist}(X,Y)]
\label{powerdecaytruncated}
\end{equation}
for the exponentially decaying interaction $h_X$, 
where ${\tilde \mu}=\mu/(1+2v/\Delta E)$. 
This proves Theorem~\ref{clusterboundbosonic}.
The corresponding bound for finite-range interactions 
was already obtained in \cite{Hastings2}. 
Using the definition (\ref{GSexpect}) of 
the expectation $\langle\cdots\rangle_{0,\Lambda}$ and the identity, 
\begin{equation}
\left\langle A_XP_0B_Y\right\rangle_{0,\Lambda}=
\left\langle B_YP_0A_X\right\rangle_{0,\Lambda},
\label{idABBA}
\end{equation}
for the integral in the right-hand side of (\ref{extractTrcorr}),  
we obtain   
\begin{eqnarray}
& &\left|\left\langle A_XB_Y\right\rangle_{0,\Lambda}
-\left\langle A_XP_{0,\Lambda}B_Y\right\rangle_{0,\Lambda}\right|\ret
&\le&{\rm Const.}\times\cases{[1+{\rm dist}(X,Y)]^{-{\tilde\eta}}, 
                             & for power-law decaying $h_X$;\cr
                    \exp[-{\tilde \mu}\>{\rm dist}(X,Y)], 
                             & for exponentially decaying $h_X$\cr}
\label{expcluster0}
\end{eqnarray}
for any finite lattice $\Lambda_s\supset X,Y$ in the same way as in the above. 

Next consider the case that the pair, $A_X,B_Y$, is fermionic. Note that 
\begin{eqnarray}
& &\left\langle\Phi_{0,\nu},\{A_X(t),B_Y\}\Phi_{0,\nu'}\right\rangle\ret
&=&\left\langle\Phi_{0,\nu}, A_X(t)(1-P_{0,\Lambda})B_Y\Phi_{0,\nu'}\right\rangle
+\left\langle\Phi_{0,\nu},B_Y(1-P_{0,\Lambda})A_X(t)\Phi_{0,\nu'}\right\rangle\ret
&+&\left\langle\Phi_{0,\nu},A_X(t)P_{0,\Lambda}B_Y\Phi_{0,\nu'}\right\rangle
+\left\langle\Phi_{0,\nu},B_YP_{0,\Lambda}A_X(t)\Phi_{0,\nu'}\right\rangle.
\label{ABanticommuexpect} 
\end{eqnarray}
Since the difference between bosonic and fermionic observables 
is in the signs of some terms, one has 
\begin{eqnarray}
& &\left|\omega(A_XB_Y)
-\frac{1}{2}\left[\omega(A_XP_0B_Y)-\omega(B_YP_0A_X)\right]\right|\ret
&\le&{\rm Const.}\times\cases{[1+{\rm dist}(X,Y)]^{-{\tilde\eta}}, 
                             & for power-law decaying $h_X$;\cr
                    \exp[-{\tilde \mu}\>{\rm dist}(X,Y)], 
                             & for exponentially decaying $h_X$.\cr}
\end{eqnarray}
In particular, thanks to the identity (\ref{idABBA}), we obtain 
\begin{eqnarray}
\left|\left\langle A_XB_Y\right\rangle_0\right|
&\le&{\rm Const.}\times\cases{[1+{\rm dist}(X,Y)]^{-{\tilde\eta}}, 
                             & for power-law decaying $h_X$;\cr
                    \exp[-{\tilde \mu}\>{\rm dist}(X,Y)], 
                             & for exponentially decaying $h_X$.\cr}\ret
\end{eqnarray}
This is nothing but the desired bound. 
We stress that, for infinite degeneracy of the infinite-volume ground 
state, this upper bound is also justified in the same argument 
with the dominated convergence theorem.

\Section{Vanishing of the matrix elements in the ground state}
\label{sec:vanishing}

The aim of this section is to prove the bound (\ref{decaycorrboson}) 
for the correlation   
and discuss an extension of Theorem~\ref{theorem:bosonic} to a system having 
infinite degeneracy of infinite-volume ground state. 
The latter result is summarized as Theorem~\ref{theorem:vanishfinite} below. 
We will give only the proof of Theorem~\ref{theorem:vanishfinite} 
because Theorem~\ref{theorem:bosonic} is proved in the same way.   
By the clustering bounds (\ref{expdecaytruncated}) and (\ref{powerdecaytruncated}), 
it is sufficient to show that all the matrix elements, 
$\left\langle\Phi_{0,\nu'}A_X\Phi_{0,\nu}\right\rangle$, in the sector 
of the ground state are vanishing. The key idea of the proof is 
to estimate the absolute values of the matrix elements 
by using the self-similarity condition and  
the decay bound (\ref{corrbound}) below of the correlations 
at a sufficiently large distance.  

We denote by $q_\Lambda$ 
the degeneracy of the sector of the ground state for the finite lattice $\Lambda$, 
and we allow $q_\Lambda\rightarrow\infty$ as $|\Lambda_s|\uparrow\infty$. 
We write $m=q_\Lambda^2$ for short. 
To begin with, we write the bound (\ref{expcluster0}) as 
\begin{equation}
\left|\left\langle A_XB_Y\right\rangle_{0,\Lambda}
-\left\langle A_XP_{0,\Lambda}B_Y\right\rangle_{0,\Lambda}\right|
\le G_0({\rm dist}(X,Y)), 
\label{clusterbound}
\end{equation}
where we have written the upper bound of the right-hand side 
by the function $G_0$ of the distance. We assume that the following bound holds:  
\begin{equation}
\left|\left\langle A_XB_Y\right\rangle_{0,\Lambda}\right|\le G_1({\rm dist}(X,Y))
\label{corrbound}
\end{equation}
with an upper bound $G_1$ which is vanishing at the infinite distance. 
Further we define ${\tilde G}_\Lambda$ as 
\begin{equation}
{\tilde G}_\Lambda(A_X,B_Y):=
\max\left\{G_0({\rm dist}(X,Y)),G_1({\rm dist}(X,Y))\right\}.
\label{tildeG}
\end{equation}

\begin{theorem}
\label{theorem:vanishfinite}
Let $\omega$ be a ground-state expectation (\ref{omega}) 
in the infinite volume limit, and let $A_X,B_Y$ be a pair of bosonic observable 
with compact supports $X,Y$. 
Assume that there exists a uniform spectral gap $\Delta E>0$ above the 
ground state sector in the spectrum of the Hamiltonian 
$H_\Lambda$ in the sense of Definition~\ref{definition}.
Suppose that, for any given $\epsilon>0$, 
there exists $M_0>0$ such that, for any large lattice $\Lambda$ 
satisfying $|\Lambda_s|\ge M_0$, there exists a set of observables, 
$B^{(j)}$, $j=1,2,\ldots,m$, and a set of transformations, $R_j$, 
$j=1,2,\ldots,m$, satisfying the following conditions: 
Any pair of the observables, $A_X, B^{(1)},\ldots,B^{(m)}$, is bosonic, 
\begin{equation}
B^{(j)}=R_j(A),\quad (B^{(j)})^\dagger=R_j(A_X^\dagger)\quad\mbox{and} 
\quad R_j(H_\Lambda)=H_\Lambda,
\label{Rassum}
\end{equation}
and 
\begin{equation}
q_\Lambda^3\mathop{\max_{i,j\in\{0,1,\ldots,m\}:}}_{i\ne j}
\left\{{\tilde G}_\Lambda\left(\left(B^{(i)}\right)^\dagger,B^{(j)}\right)\right\}
<\epsilon,
\label{tildeGbound}
\end{equation}
where we have written $B^{(0)}=A_X^\dagger$. Then we have the bound, 
\begin{equation}
\left|\omega(A_XB_Y)\right|
\le{\rm Const.}\times\cases{[1+{\rm dist}(X,Y)]^{-{\tilde\eta}}, 
                             & for power-law decaying $h_X$;\cr
                    \exp[-{\tilde \mu}\>{\rm dist}(X,Y)], 
                             & for exponentially decaying $h_X$,\cr}
\end{equation}
in the infinite volume limit. 
\end{theorem}

\begin{proof}{Proof}
{From} the bound (\ref{expdecaytruncated}) or (\ref{powerdecaytruncated}) and 
the Schwarz inequality,  
\begin{equation}
\left|\omega(A_XP_0B_Y)\right|^2\le
\omega(A_XP_0A_X^\dagger)\omega(B_Y^\dagger P_0B_Y),
\end{equation}
it is sufficient to show $\omega(A_XP_0A_X^\dagger)=\omega(A_X^\dagger P_0A_X)=0$. 
Further, we have 
\begin{equation}
\left\langle\Phi,AP_{0,\Lambda}A^\dagger\Phi\right\rangle\le q_\Lambda
\left\langle AP_{0,\Lambda}A^\dagger\right\rangle_{0,\Lambda}=
q_\Lambda
\left\langle A^\dagger P_{0,\Lambda}A\right\rangle_{0,\Lambda}
\end{equation}
for any ground state vector $\Phi$ with norm one and any observable $A$ on the finite 
lattice $\Lambda$. Therefore we estimate 
$q_\Lambda\left\langle A_XP_{0,\Lambda}A_X^\dagger\right\rangle_{0,\Lambda}$. 

Note that, from the clustering bound (\ref{clusterbound}), (\ref{corrbound}) 
and (\ref{tildeG}), we have 
\begin{eqnarray}
\left|\left\langle A_XP_{0,\Lambda}B_Y\right\rangle_{0,\Lambda}\right|
&\le&\left|\left\langle A_XB_Y\right\rangle_{0,\Lambda}\right|
+\left|\left\langle A_XB_Y\right\rangle_{0,\Lambda}
-\left\langle A_XP_{0,\Lambda}B_Y\right\rangle_{0,\Lambda}\right|\ret
&\le&2{\tilde G}_\Lambda(A_X,B_Y).
\label{APBbound} 
\end{eqnarray}
We define
\begin{equation}
B_i^{(j)}:=
\left\langle \Phi_{0,\nu'},B^{(j)}\Phi_{0,\nu}\right\rangle
\end{equation}
for $j=0,1,\ldots,m$ and for the finite lattice $\Lambda$, where  
we have written $i=(\nu',\nu)$ with $i=1,2,\ldots,m$ for short.
Since $(B_1^{(j)},B_2^{(j)},\ldots,B_m^{(j)})$ is an $m$-dimensional vector,   
there exist complex numbers, $C_j, j=0,1,\ldots,m$, 
such that, at least, one of $C_j$ is nonvanishing and that  
\begin{equation}
\sum_{j=0}^m C_jB_i^{(j)}=0. 
\end{equation}
Let $\ell$ be the index which satisfies $|C_\ell|=\max\{|C_0|,|C_1|,\ldots,|C_m|\}$. 
Clearly, we have 
\begin{equation}
B_i^{(\ell)}=-\sum_{j\ne \ell}\frac{C_j}{C_\ell}B_i^{(j)}. 
\end{equation}
Therefore
\begin{eqnarray}
\left\langle (B^{(\ell)})^\dagger P_{0,\Lambda}B^{(\ell)}\right\rangle_{0,\Lambda}
=\frac{1}{q_\Lambda}\sum_{i=1}^m \left|B_i^{(\ell)}\right|^2
&=&-\sum_{j\ne \ell}\frac{C_j}{C_\ell}
\frac{1}{q_\Lambda}\sum_{i=1}^m(B_i^{(\ell)})^\ast B_i^{(j)}\ret
&\le&m\max_{j\ne \ell}
\left\{\left|\left\langle (B^{(\ell)})^\dagger P_{0,\Lambda}
B^{(j)}\right\rangle_{0,\Lambda}\right|\right\}\ret
&\le&2q_\Lambda^2
\max_{j\ne \ell}\left\{
{\tilde G}_\Lambda\left((B^{(\ell)})^\dagger,B^{(j)}\right)
\right\},
\end{eqnarray}
where we have used the inequality (\ref{APBbound}) for getting 
the last bound. 
When $\ell=0$, we obtain 
\begin{equation}
q_\Lambda\left\langle A_XP_{0,\Lambda}A_X^\dagger\right\rangle_{0,\Lambda}
\le 2\epsilon
\end{equation}
from $B^{(0)}=A_X^\dagger$ and the assumption (\ref{tildeGbound}). 
When $\ell\ne 0$, we reach the same conclusion by using 
the relation, 
\begin{equation}
\left\langle A_X^\dagger P_{0,\Lambda}A_X\right\rangle_{0,\Lambda}
=\left\langle R_\ell(A_X^\dagger)P_{0,\Lambda}R_\ell(A)\right\rangle_{0,\Lambda}=
\left\langle \left(B^{(\ell)}\right)^\dagger P_{0,\Lambda}
B^{(\ell)}\right\rangle_{0,\Lambda}, 
\label{expectAdaggerA} 
\end{equation}
which is derived from the assumption (\ref{Rassum}).  
\end{proof}

\noindent
{\bf Remark:} 
1. The advantage of Theorem~\ref{theorem:vanishfinite} is that it is easier to 
find $B^{(j)}$ and $R_j$ because of the finiteness of the lattice. 
Actually one can construct $B^{(j)}$, $R_j$ and $\Lambda$ 
satisfying the requirement by connecting $m$ copies of a small, finite lattice 
to each other at their boundaries. 
But, if the degeneracy $q_\Lambda$ exceeds 
$\sqrt{|\Lambda_s|}$, we cannot find the observables, $B^{(j)}$, 
and the transformations, $R_j$. Therefore our argument does not work in 
such cases.
\smallskip

\noindent
2. Under the weaker assumption, 
\begin{equation}
q_\Lambda^2\mathop{\max_{i,j\in\{0,1,\ldots,m\}:}}_{i\ne j}
\left\{{\tilde G}_\Lambda\left(\left(B^{(i)}\right)^\dagger,B^{(j)}\right)\right\}
<\epsilon,
\end{equation}
than (\ref{tildeGbound}), we can obtain the bound, 
\begin{equation}
\left|\left\langle A_XB_Y\right\rangle_0\right|
\le{\rm Const.}\times\cases{[1+{\rm dist}(X,Y)]^{-{\tilde\eta}}, 
                             & for power-law decaying $h_X$;\cr
                    \exp[-{\tilde \mu}\>{\rm dist}(X,Y)], 
                             & for exponentially decaying $h_X$,\cr}
\end{equation}
in the infinite volume limit.
\smallskip

\noindent
3. Consider the situation of the above Remark~2 or the case with a finite 
degeneracy of the infinite-volume ground state. Then, 
instead of introducing the transformations $R_j$, we can directly require 
\begin{equation}
\left\langle A_X^\dagger P_0A_X\right\rangle_0
=\left\langle \left(B^{(j)}\right)^\dagger P_0B^{(j)}\right\rangle_0\quad\mbox{for }
j=1,2,\ldots,m,
\end{equation}
in the infinite volume limit, and at infinite distance between the observables $A_X$ 
and $B^{(j)}$.
\smallskip

\appendix

\Section{Lieb-Robinson bound for group velocity}
\label{LRbound}

Quite recently, Nachtergaele and Sims \cite{NachSim} have extended 
the Lieb-Robinson bound \cite{LR} 
to a wide class of models with long-range, exponentially decaying interactions.  
In this appendix, we further extend the bound to the power-law 
decaying interactions. We also tighten the bound on the exponentially 
decaying case. (See Assumption~\ref{assumption:expodecaysum} compared to 
that in \cite{NachSim}).  
However, in our proof, the time $t$ must be real.

In the following, we treat only the case with bosonic observables 
and with the power-law decaying interaction $h_X$ 
because the other cases including the previous results can be treated in the same way.  

\begin{theorem}
\label{LRpowerdecay}
Let $A_X,B_Y$ be a pair of bosonic observables 
with the compact support, $X,Y$, respectively.  
Assume that the system satisfies the conditions in 
Assumption~\ref{assumption:powerdecay} or \ref{assumption:powerdecaysum}. Then  
\begin{equation}
\Vert [A_X(t),B_Y]\Vert\le C\Vert A_X\Vert \Vert B_Y\Vert|X||Y|
\frac{e^{v|t|}-1}{[1+{\rm dist}(X,Y)]^\eta}\quad\mbox{for }\ 
{\rm dist}(X,Y)>0,
\label{LRboundpower} 
\end{equation}
where the positive constants, $C$ and $v$, depend only on the interaction of the 
Hamiltonian and the metric of the lattice. 
\end{theorem}

\noindent
{\bf Remark:} The same bound for fermionic observables is 
obtained by replacing the commutator with the anticommutator in 
the left-hand side. 
\bigskip

For exponentially decaying interaction $h_X$, the following 
bound is valid: 

\begin{theorem}
Let $A_X,B_Y$ be a pair of bosonic observables with 
the compact support, $X,Y$, respectively.  
Assume that the system satisfies the conditions in 
Assumption~\ref{assumption:expodecay} or \ref{assumption:expodecaysum}. Then  
\begin{equation}
\Vert [A_X(t),B_Y]\Vert\le C\Vert A_X\Vert \Vert B_Y\Vert|X||Y|
\exp[-\mu\>{\rm dist}(X,Y)]\left[e^{v|t|}-1\right]\quad\mbox{for }\ 
{\rm dist}(X,Y)>0, 
\end{equation}
where the positive constants, $C$ and $v$, depend only on the interaction of the 
Hamiltonian and the metric of the lattice. 
\end{theorem}

\noindent
{\bf Remark:} For the proof under Assumption~\ref{assumption:expodecay}, 
we rely on the inequalities, (\ref{expepspoly}) and (\ref{expotwoprod}). 
Assumption~\ref{assumption:expodecaysum} is milder than 
that in ref.~\cite{NachSim} as remarked in Section~\ref{Sec:main}. 
\bigskip

We assume that the volume $|\Lambda_s|$ of the lattice 
$\Lambda$ is finite. If it is necessary to consider the infinite volume limit,  
we take the limit after deriving the desired Lieb-Robinson bounds which 
hold uniformly in the size of the lattice. 
Let $A,B$ be observables supported by compact sets, $X,Y\subset \Lambda_s$, 
respectively. The time evolution of $A$ is given by 
$A(t)=e^{itH_\Lambda}Ae^{-itH_\Lambda}$.
First, let us derive the inequality (\ref{Commnormbound}) below 
for the commutator $[A(t),B]$. 
We assume $t>0$ because the negative $t$ can be treated in the same way. 
Let $\epsilon=t/N$ with a large positive integer $N$, and let 
\begin{equation}
t_n=\frac{t}{N}n\quad\mbox{for}\ n=0,1,\ldots, N. 
\end{equation}
Then we have 
\begin{equation}
\left\Vert[A(t),B]\right\Vert-\left\Vert[A(0),B]\right\Vert 
=\sum_{i=0}^{N-1}\epsilon\times
\frac{\left\Vert[A(t_{n+1}),B]\right\Vert-\left\Vert[A(t_n),B]\right\Vert}{\epsilon}.
\label{sumid} 
\end{equation}
In order to obtain the bound (\ref{Commnormbound}) below, 
we want to estimate the summand in the right-hand side. 
To begin with, we note that the identity,  
$\left\Vert U^\ast OU\right\Vert=\Vert O\Vert$, holds for any observable $O$ 
and for any unitary operator $U$. Using this fact, we have  
\begin{eqnarray}
\left\Vert[A(t_{n+1}),B]\right\Vert-\left\Vert[A(t_n),B]\right\Vert
&=&\left\Vert[A(\epsilon),B(-t_n)]\right\Vert-\left\Vert[A,B(-t_n)]\right\Vert\ret
&\le&\left\Vert[A+i\epsilon[H_\Lambda,A],B(-t_n)]\right\Vert
-\left\Vert[A,B(-t_n)]\right\Vert+{\cal O}(\epsilon^2)\ret
&=&\left\Vert[A+i\epsilon[I_X,A],B(-t_n)]\right\Vert
-\left\Vert[A,B(-t_n)]\right\Vert+{\cal O}(\epsilon^2)\ret
\label{difnorm}
\end{eqnarray}
with 
\begin{equation}
I_X=\sum_{Z:Z\cap X\ne \emptyset}h_Z,
\label{defIX}
\end{equation}
where we have used 
\begin{equation}
A(\epsilon)=A+i\epsilon[H_\Lambda,A]+{\cal O}(\epsilon^2)
\end{equation}
and the triangle inequality. Further, by using 
\begin{equation}
A+i\epsilon[I_X,A]=e^{i\epsilon I_X}Ae^{-i\epsilon I_X}+{\cal O}(\epsilon^2), 
\end{equation}
we have 
\begin{eqnarray}
\left\Vert[A+i\epsilon[I_X,A],B(-t_n)]\right\Vert
&\le&\left\Vert[e^{i\epsilon I_X}Ae^{-i\epsilon I_X},B(-t_n)]\right\Vert
+{\cal O}(\epsilon^2)\ret
&=&\left\Vert[A,e^{-i\epsilon I_X}B(-t_n)e^{i\epsilon I_X}]\right\Vert
+{\cal O}(\epsilon^2)\ret
&\le&\left\Vert[A,B(-t_i)-i\epsilon[I_X,B(-t_n)]]\right\Vert
+{\cal O}(\epsilon^2)\ret
&\le&\left\Vert[A,B(-t_n)]\right\Vert+\epsilon\left\Vert[A,[I_X,B(-t_n)]]\right\Vert
+{\cal O}(\epsilon^2). 
\end{eqnarray}
Substituting this into the right-hand side in the last line of (\ref{difnorm}), 
we obtain 
\begin{eqnarray}
\left\Vert[A(t_{n+1}),B]\right\Vert-\left\Vert[A(t_n),B]\right\Vert
&\le&\epsilon\left\Vert[A,[I_X,B(-t_n)]]\right\Vert
+{\cal O}(\epsilon^2)\ret
&\le&2\epsilon\Vert A\Vert\left\Vert[I_X(t_n),B]\right\Vert
+{\cal O}(\epsilon^2).
\end{eqnarray}
Further, substituting this into the right-hand side of (\ref{sumid}) and  
using (\ref{defIX}), we have 
\begin{eqnarray}
\left\Vert[A(t),B]\right\Vert-\left\Vert[A(0),B]\right\Vert 
&\le&2\Vert A\Vert\sum_{n=0}^{N-1}\epsilon\times\left\Vert[I_X(t_n),B]\right\Vert
+{\cal O}(\epsilon)\ret
&\le&2\Vert A\Vert\sum_{Z:Z\cap X\ne \emptyset}\sum_{n=0}^{N-1}\epsilon\times
\left\Vert[h_Z(t_n),B]\right\Vert+{\cal O}(\epsilon). 
\end{eqnarray}
Since $h_Z(t)$ is the continuous function of the time $t$ for a finite volume, 
the sum in the right-hand side converges to the integral in 
the limit $\epsilon\downarrow 0$ ($N\uparrow\infty$) 
for any fixed finite lattice $\Lambda$. In consequence, we obtain
\begin{equation}
\left\Vert[A(t),B]\right\Vert-\left\Vert[A(0),B]\right\Vert 
\le 2\Vert A\Vert\sum_{Z:Z\cap X\ne \emptyset}\int_0^{|t|}ds 
\left\Vert[h_Z(s),B]\right\Vert.
\label{Commnormbound} 
\end{equation}

We define 
\begin{equation}
C_B(X,t):=\sup_{A\in{\cal A}_X}\frac{\Vert[A(t),B]\Vert}{\Vert A\Vert},
\label{CBXt}
\end{equation}
where ${\cal A}_X$ is the set of observables supported by 
the compact set $X$. Then we have\footnote{Since the local interaction $h_Z$ 
with $Z\subset X$ does not change the support $X$ of $A$ in 
the time evolution, we can expect that the sum in the right-hand side 
of (\ref{CBXtbound}) can be restricted to the set $Z$ 
satisfying $Z\cap X \neq \emptyset$ and $Z\setminus X\neq\emptyset$. 
However, this restriction does not affect the resulting Lieb-Robinson 
bound. Therefore we omit the discussion.} 
\begin{equation}
C_B(X,t)\le C_B(X,0)+2\sum_{Z:Z\cap X\ne \emptyset}\Vert h_Z\Vert
\int_0^{|t|}ds\> C_B(Z,s)
\label{CBXtbound}
\end{equation}
from the above bound (\ref{Commnormbound}).

We recall that the observables, $A$ and $B$, are, respectively, 
supported by the compact sets, $X,Y\subset \Lambda_s$. 
Assume ${\rm dist}(X,Y)>0$. 
Then we have $C_B(X,0)=0$ from the definition of $C_B(X,t)$, and note that 
\begin{equation}
C_B(Z,0)\le\cases{2\Vert B\Vert, & for $Z\cap Y\ne\emptyset$;\cr
                          0, & otherwise.\cr}
\end{equation}
Using these facts and the above bound (\ref{CBXtbound}) iteratively, 
we obtain 
\begin{eqnarray}
C_B(X,t)&\le&2\sum_{Z_1:Z_1\cap X\ne\emptyset}\Vert h_{Z_1}\Vert
\int_0^{|t|}ds_1\> C_B(Z_1,s_1)\ret
&\le&2\sum_{Z_1:Z_1\cap X\ne\emptyset}\Vert h_{Z_1}\Vert
\int_0^{|t|}ds_1\> C_B(Z_1,0)\ret
&+&2^2\sum_{Z_1:Z_1\cap X\ne\emptyset}\Vert h_{Z_1}\Vert
\sum_{Z_2:Z_2\cap Z_1\ne\emptyset}\Vert h_{Z_2}\Vert
\int_0^{|t|}ds_1\int_0^{|s_1|}ds_2\> C_B(Z_2,s_2)\ret
&\le&2\Vert B\Vert(2|t|)\sum_{Z_1:Z_1\cap X\ne\emptyset,Z_1\cap Y\ne\emptyset}
\Vert h_{Z_1}\Vert\ret
&+&2\Vert B\Vert\frac{(2|t|)^2}{2!}
\sum_{Z_1:Z_1\cap X\ne\emptyset}\Vert h_{Z_1}\Vert
\sum_{Z_2:Z_2\cap Z_1\ne\emptyset,Z_2\cap Y\ne\emptyset}\Vert h_{Z_2}\Vert\ret
&+&2\Vert B\Vert\frac{(2|t|)^3}{3!}
\sum_{Z_1:Z_1\cap X\ne\emptyset}\Vert h_{Z_1}\Vert
\sum_{Z_2:Z_2\cap Z_1\ne\emptyset}\Vert h_{Z_2}\Vert
\sum_{Z_3:Z_3\cap Z_2\ne\emptyset,Z_3\cap Y\ne\emptyset}\Vert h_{Z_3}\Vert+\cdots
\ret
\label{CBexpabound}
\end{eqnarray}
\bigskip

\begin{proof}{Proof of Theorem~\ref{LRpowerdecay} under 
Assumption~\ref{assumption:powerdecay}}
The first sum in the power series (\ref{CBexpabound}) is estimated as 
\begin{eqnarray}
\sum_{Z_1:Z_1\cap X\ne\emptyset,Z_1\cap Y\ne\emptyset}
\Vert h_{Z_1}\Vert
&\le&\sum_{x\in X}\sum_{y\in Y}\sum_{Z_1\ni x,y}\Vert h_{Z_1}\Vert\ret
&\le&\frac{\lambda_0|X||Y|}{[1+{\rm dist}(X,Y)]^\eta} 
\end{eqnarray}
from the assumption (\ref{powerdecayhX}). The second, double sum is estimated as 
\begin{eqnarray}
& &\sum_{Z_1:Z_1\cap X\ne\emptyset}\Vert h_{Z_1}\Vert
\sum_{Z_2:Z_2\cap Z_1\ne\emptyset,Z_2\cap Y\ne\emptyset}\Vert h_{Z_2}\Vert\ret
&\le&\sum_{x\in X}\sum_{y\in Y}\sum_{z_{12}\in\Lambda_s}\>
\sum_{Z_1\ni x,z_{12}}\Vert h_{Z_1}\Vert
\sum_{Z_2\ni z_{12},y}\Vert h_{Z_2}\Vert\ret
&\le&\sum_{x\in X}\sum_{y\in Y}\sum_{z_{12}\in\Lambda_s}
\frac{\lambda_0}{[1+{\rm dist}(x,z_{12})]^\eta}
\frac{\lambda_0}{[1+{\rm dist}(z_{12},y)]^\eta}\ret
&\le&\frac{\lambda_0^2p_0|X||Y|}{[1+{\rm dist}(X,Y)]^\eta},
\end{eqnarray}
where we have used the assumptions (\ref{powerdecayhX}) 
and (\ref{Propaineq}). Similarly, the third, triple sum can be estimated 
as 
\begin{eqnarray}
& &\sum_{Z_1:Z_1\cap X\ne\emptyset}\Vert h_{Z_1}\Vert
\sum_{Z_2:Z_2\cap Z_1\ne\emptyset}\Vert h_{Z_2}\Vert
\sum_{Z_3:Z_3\cap Z_2\ne\emptyset,Z_3\cap Y\ne\emptyset}\Vert h_{Z_3}\Vert\ret
&\le&\sum_{x\in X}\sum_{y\in Y}\sum_{z_{12}\in\Lambda_s}\sum_{z_{23}\in\Lambda_s}\>
\sum_{Z_1\ni x,z_{12}}\Vert h_{Z_1}\Vert
\sum_{Z_2\ni z_{12},z_{23}}\Vert h_{Z_2}\Vert
\sum_{Z_3\ni z_{23},y}\Vert h_{Z_3}\Vert\ret
&\le&\sum_{x\in X}\sum_{y\in Y}\sum_{z_{12}\in\Lambda_s}\sum_{z_{23}\in\Lambda_s}
\frac{\lambda_0}{[1+{\rm dist}(x,z_{12})]^\eta}
\frac{\lambda_0}{[1+{\rm dist}(z_{12},z_{23})]^\eta}
\frac{\lambda_0}{[1+{\rm dist}(z_{23},y)]^\eta}\ret
&\le&\sum_{x\in X}\sum_{y\in Y}\sum_{z_{12}\in\Lambda_s}
\frac{\lambda_0}{[1+{\rm dist}(x,z_{12})]^\eta}
\frac{\lambda_0^2p_0}{[1+{\rm dist}(z_{12},y)]^\eta}\ret
&\le&\frac{\lambda_0^3p_0^2|X||Y|}{[1+{\rm dist}(X,Y)]^\eta}. 
\end{eqnarray}
{From} these observations, we have 
\begin{eqnarray}
C_B(X,t)&\le&\frac{2\Vert B\Vert |X||Y|}{[1+{\rm dist}(X,Y)]^\eta}
\left\{2|t|\lambda_0+\frac{(2|t|)^2}{2!}\lambda_0^2p_0
+\frac{(2|t|)^3}{3!}\lambda_0^3p_0^2+\cdots\right\}\ret
&=&\frac{2p_0^{-1}\Vert B\Vert |X||Y|}{[1+{\rm dist}(X,Y)]^\eta}
\left\{\exp[2\lambda_0p_0|t|]-1\right\}. 
\end{eqnarray}
Consequently, we obtain 
\begin{equation}
\Vert[A(t),B]\Vert\le\frac{2p_0^{-1}\Vert A\Vert
\Vert B\Vert |X||Y|}{[1+{\rm dist}(X,Y)]^\eta}
\left\{\exp[2\lambda_0p_0|t|]-1\right\} 
\end{equation}
from (\ref{CBXt}). 
\end{proof}

\begin{proof}{Proof of Theorem~\ref{LRpowerdecay} under 
Assumption~\ref{assumption:powerdecaysum}}
The first sum in the power series (\ref{CBexpabound}) is estimated as 
\begin{eqnarray}
\sum_{Z_1:Z_1\cap X\ne\emptyset,Z_1\cap Y\ne\emptyset}
\Vert h_{Z_1}\Vert
&\le&\sum_{x\in X}\sum_{y\in Y}\sum_{Z_1\ni x,y}\Vert h_{Z_1}\Vert\ret
&\le&\sum_{x\in X}\sum_{y\in Y}\sum_{Z_1\ni x,y}\Vert h_{Z_1}\Vert
[1+{\rm dist}(x,y)]^{-\eta}[1+{\rm diam}(Z_1)]^\eta\ret
&\le&[1+{\rm dist}(X,Y)]^{-\eta}|X||Y|s_0, 
\end{eqnarray}
where 
\begin{equation}
s_0=\sup_x\sum_{Z\ni x}\Vert h_Z\Vert[1+{\rm diam}(Z)]^\eta.
\end{equation}
Clearly, this constant $s_0$ is finite from the assumption (\ref{s1bound}).
The second, double sum is estimated as 
\begin{eqnarray}
& &\sum_{Z_1:Z_1\cap X\ne\emptyset}\Vert h_{Z_1}\Vert
\sum_{Z_2:Z_2\cap Z_1\ne\emptyset,Z_2\cap Y\ne\emptyset}\Vert h_{Z_2}\Vert\ret
&\le&\sum_{x\in X}\sum_{y\in Y}\sum_{z_{12}\in\Lambda_s}
\sum_{Z_1\ni x,z_{12}}\Vert h_{Z_1}\Vert
\sum_{Z_2\ni z_{12},y}\Vert h_{Z_2}\Vert\ret
&\le&\sum_{x\in X}\sum_{y\in Y}\sum_{z_{12}\in\Lambda_s}
[1+{\rm dist}(x,z_{12})]^{-\eta}[1+{\rm dist}(z_{12},y)]^{-\eta}\ret
&\times&\sum_{Z_1\ni x,z_{12}}\Vert h_{Z_1}\Vert[1+{\rm diam}(Z_1)]^\eta
\sum_{Z_2\ni z_{12},y}\Vert h_{Z_2}\Vert[1+{\rm diam}(Z_2)]^\eta\ret
&\le&[1+{\rm dist}(X,Y)]^{-\eta}
\sum_{x\in X}\sum_{y\in Y}\sum_{z_{12}\in\Lambda_s}
\sum_{Z_1\ni x,z_{12}}\Vert h_{Z_1}\Vert[1+{\rm diam}(Z_1)]^\eta\ret
&\times&\sum_{Z_2\ni z_{12},y}\Vert h_{Z_2}\Vert[1+{\rm diam}(Z_2)]^\eta\ret
&\le&[1+{\rm dist}(X,Y)]^{-\eta}
\sum_{x\in X}\sum_{y\in Y}\sum_{Z_1\ni x}\Vert h_{Z_1}\Vert[1+{\rm diam}(Z_1)]^\eta\ret
&\times&
\sum_{Z_2\ni y}\Vert h_{Z_2}\Vert|Z_2|[1+{\rm diam}(Z_2)]^\eta\ret
&\le&[1+{\rm dist}(X,Y)]^{-\eta}|X||Y|s_0s_1,
\end{eqnarray}
where we have used the assumption (\ref{s1bound}) and the inequality, 
\begin{eqnarray}
& &[1+{\rm dist}(x,z)]^{-\eta}[1+{\rm dist}(z,y)]^{-\eta}\ret
&=&[1+{\rm dist}(x,z)+{\rm dist}(z,y)+{\rm dist}(x,z)
{\rm dist}(z,y)]^{-\eta}\ret
&\le&[1+{\rm dist}(x,z)+{\rm dist}(z,y)]^{-\eta}\ret
&\le&[1+{\rm dist}(x,y)]^{-\eta}, 
\end{eqnarray} 
for any $z\in\Lambda_s$. 
Similarly, the third, triple sum can be estimated as 
\begin{eqnarray}
& &\sum_{Z_1:Z_1\cap X\ne\emptyset}\Vert h_{Z_1}\Vert
\sum_{Z_2:Z_2\cap Z_1\ne\emptyset}\Vert h_{Z_2}\Vert
\sum_{Z_3:Z_3\cap Z_2\ne\emptyset,Z_3\cap Y\ne\emptyset}\Vert h_{Z_3}\Vert\ret
&\le&\sum_{x\in X}\sum_{y\in Y}\sum_{z_{12}\in\Lambda_s}\sum_{z_{23}\in\Lambda_s}\>
\sum_{Z_1\ni x,z_{12}}\Vert h_{Z_1}\Vert
\sum_{Z_2\ni z_{12},z_{23}}\Vert h_{Z_2}\Vert
\sum_{Z_3\ni z_{23},y}\Vert h_{Z_3}\Vert\ret
&\le&[1+{\rm dist}(X,Y)]^{-\eta}
\sum_{x\in X}\sum_{y\in Y}\sum_{z_{12}\in\Lambda_s}\sum_{z_{23}\in\Lambda_s}\>
\sum_{Z_1\ni x,z_{12}}\Vert h_{Z_1}\Vert[1+{\rm diam}(Z_1)]^\eta\ret
&\times&\sum_{Z_2\ni z_{12},z_{23}}\Vert h_{Z_2}\Vert
[1+{\rm diam}(Z_2)]^\eta
\sum_{Z_3\ni z_{23},y}\Vert h_{Z_3}\Vert
[1+{\rm diam}(Z_3)]^\eta\ret
&\le&[1+{\rm dist}(X,Y)]^{-\eta}|X||Y|s_0s_1^2.
\end{eqnarray}
{From} these observations, we have 
\begin{eqnarray}
C_B(X,t)&\le&\frac{2s_0s_1^{-1}\Vert B\Vert|X||Y|}{[1+{\rm dist}(X,Y)]^\eta}
\sum_{n=1}^\infty\frac{(2s_1|t|)^n}{n!}\ret
&=&\frac{2s_0s_1^{-1}\Vert B\Vert|X||Y|}{[1+{\rm dist}(X,Y)]^\eta}
\left\{\exp[2s_1|t|]-1\right\}.
\end{eqnarray}
As a result, we obtain
\begin{equation}
\Vert[A(t),B]\Vert\le\frac{2s_0s_1^{-1}\Vert A\Vert
\Vert B\Vert|X||Y|}{[1+{\rm dist}(X,Y)]^\eta}
\left\{\exp[2s_1|t|]-1\right\} 
\end{equation}
from (\ref{CBXt}).
\end{proof}

\bigskip\bigskip

\noindent
{\bf Acknowledgements:} We would like to thank 
Jens Eisert and Tobias Osborne for useful comments.
TK thanks Bruno Nachtergaele and Hal Tasaki for helpful discussions. 
MBH was supported by US DOE W-7405-ENG-36.
\bigskip\bigskip


\end{document}